# Capacity-Achievability of Polar Codes and Precoded Polar Codes under List Decoding: They are Good both Practically and Theoretically

Zhuo Li, Lijuan Xing, and Ba-Zhong Shen

*Abstract*—**Polar codes under successive cancellation decoding proposed by Arıkan provably achieve the symmetric capacity of any given binary-input discrete memoryless channel. The successive cancellation list decoder for polar codes was described by Tal and Vardy as a generalization of the successive cancellation decoder of Arıkan. The performance of the successive cancellation list decoder is encouraging in practice. In this paper, we formalize the successive cancellation list decoder in our notation and prove that polar codes under successive cancellation list decoding achieve the symmetric capacity of any given binary-input discrete memoryless channel in theory as well. We also formalize the polar codes with CRC precoding of Tal and Vardy. In fact, we propose a family of more general codes, namely, precoded polar codes and prove that precoded polar codes under successive cancellation list decoding can achieve the symmetric capacity of any given binary-input discrete memoryless channel under some conditions.**

## I. INTRODUCTION AND PRELIMINARIES

LET $\mathcal{X}$ and $\mathcal{Y}$ be finite sets. A discrete channel with input alphabet $\mathcal{X}$ and output alphabet $\mathcal{Y}$ is defined as $W: \mathcal{X} \to \mathcal{Y}$ with transition probabilities $W(y|x), x \in \mathcal{X}, y \in \mathcal{Y}$.

A code for channels with input alphabet $\mathcal{X}$ and output alphabet $\mathcal{Y}$ is a pair of mappings $(f, \varphi)$, where $f$ maps some finite set $\mathcal{M} = \mathcal{M}_f$ into $\mathcal{X}$ and $\varphi$ maps $\mathcal{Y}$ into $\widehat{\mathcal{M}} \supseteq \mathcal{M}$. The elements of $\mathcal{M}$ are called messages, the mapping $f$ is the encoder and $\varphi$ is the decoder. The images of the messages under $f$ are called codewords.

Given a channel $W: \mathcal{X} \to \mathcal{Y}$, a code for channel $W$ is any code $(f, \varphi)$ as above. The probability of erroneous transmission of message $m$ is

$$e_m = e_m(W, f, \varphi) \triangleq 1 - W(\varphi^{-1}(m)|f(m))$$

where

This work was supported in part by the National Natural Science Foundation of China under Grant 61372072, in part by the 111 Project under Grant B08038, and in part by the Fundamental Research Funds for the Central Universities.
The authors are with the State Key Laboratory of Integrated Services Networks, Xidian University, Xi'an, Shannxi 710071, China (e-mail: lizhuo@xidian.edu.cn; ljxing@mail.xidian.edu.cn; bzshen@xidian.edu.cn).

$$\varphi^{-1}(m) \triangleq \{y \in \mathcal{Y}: \varphi(y) = m\}$$

and

$$W(\varphi^{-1}(m)|f(m)) \triangleq \sum_{y \in \varphi^{-1}(m)} W(y|f(m)).$$

The maximum probability of error of the code $(f, \varphi)$ is

$$e = e(W, f, \varphi) \triangleq \max_{m \in \mathcal{M}} e_m.$$

The average probability of error of the code $(f, \varphi)$ is

$$\bar{e} = \bar{e}(W, f, \varphi) \triangleq \frac{1}{|\mathcal{M}|} \sum_{m \in \mathcal{M}} e_m.$$

A channel $W: \mathcal{X} \to \mathcal{Y}$ is called a binary-input discrete memoryless channel (B-DMC) if the input alphabet $\mathcal{X} = \{0,1\}$ and the $n$ uses of the channel $W^n: \mathcal{X}^n \to \mathcal{Y}^n$ has transition probabilities

$$W^n(y_1^n|x_1^n) = \prod_{i=1}^{n} W(y_i|x_i).$$

This B-DMC is denoted by $\{W: \mathcal{X} \to \mathcal{Y}\}$ or simply $\{W\}$. An $n$-length block code for a B-DMC $\{W\}$ is a code $(f, \varphi)$ for the channel $W^n$. The rate of such a code is $\frac{1}{n}\log|\mathcal{M}_f|$.

In this paper, the base of the logarithm is assumed to be 2 unless otherwise stated. We use the notation $a_1^N$ as shorthand for denoting a row vector $(a_1, \ldots, a_N)$. Given such a vector $a_1^N$, we write $a_i^j$, $1 \le i, j \le N$, to denote the subvector $(a_i, \ldots, a_j)$; if $i > j$, $a_i^j$ is regarded as void. Given $a_1^N$ and $\mathcal{A} \subseteq \{1, \ldots, N\}$, we write $a_{\mathcal{A}}$ to denote the subvector $(a_i: i \in \mathcal{A})$; if $\mathcal{A} = \emptyset$, $a_{\mathcal{A}}$ is regarded as void. The notation $0_1^N$ is used to denote the all-zero vector. We write $|\mathcal{A}|$ to denote the number of elements in a set $\mathcal{A}$. Unless specified otherwise, all vectors, matrices, and operations on them in this paper will be over the binary field.

Given a binary-input channel $W$, the symmetric capacity of $W$ is

$$I(W) \triangleq \sum_{y \in \mathcal{Y}} \sum_{x \in \mathcal{X}} \frac{1}{2} W(y|x) \log \frac{W(y|x)}{\frac{1}{2}W(y|0) + \frac{1}{2}W(y|1)}$$

and the Bhattacharyya parameter of $W$ is

$$Z(W) \triangleq \sum_{y \in \mathcal{Y}} \sqrt{W(y|0)W(y|1)}.$$

Channel polarization is a method proposed by Arıkan [1] to construct code sequences that achieve the symmetric capacity $I(W)$ of any given B-DMC $\{W\}$. Channel polarization is an operation by which one manufactures out of $N$ uses of a B-DMC $\{W\}$ a second set of $N$ channels $\left\{W_N^{(i)}: 1 \leq i \leq N\right\}$ that show a polarization effect in the sense that, as $N$ becomes large, the symmetric capacity terms $\left\{I(W_N^{(i)})\right\}$ tend towards 0 or 1 for all but a vanishing fraction of indices $i$. This operation consists of a channel combining phase and a channel splitting phase.

We call $G_N \triangleq B_N F^{\otimes n}$ the generator matrix of order $N$ for any $N = 2^n, n \geq 0$, where $B_N$ is a permutation matrix known as bit-reversal and $F^{\otimes n}$ is the Kronecker power of $F \triangleq \begin{bmatrix} 1 & 0 \\ 1 & 1 \end{bmatrix}$ with the convention that $F^{\otimes 0} \triangleq [1]$. Then the channel combining phase combines $N$ uses of a given B-DMC $\{W\}$ in a recursive manner to produce a channel $W_N: \mathcal{X}^N \to \mathcal{Y}^N$ defined by the transition probabilities

$$W_N(y_1^N|m_1^N) \triangleq W^N(y_1^N|m_1^N G_N)$$

for all $y_1^N \in \mathcal{Y}^N, m_1^N \in \mathcal{X}^N$.

Having synthesized the channel $W_N$ out of $W^N$, the channel splitting phase is to split $W_N$ back into a set of $N$ binary-input coordinate channels $W_N^{(i)}: \mathcal{X} \to \mathcal{Y}^N \times \mathcal{X}^{i-1}, 1 \leq i \leq N$, defined by the transition probabilities

$$W_N^{(i)}(y_1^N, m_1^{i-1}|m_i) \triangleq \sum_{m_{i+1}^N \in \mathcal{X}^{N-i}} \frac{1}{2^{N-1}} W_N(y_1^N|m_1^N)$$

where $(y_1^N, m_1^{i-1})$ denotes the output of $W_N^{(i)}$ and $m_i$ its input.

Then we have the following two results about channel polarization and the rate of polarization.

*Theorem 1* [1]: For any B-DMC $\{W\}$, the channels $W_N^{(i)}$ polarize in the sense that, for any fixed $\delta \in (0,1)$, as $N$ goes to infinity through powers of two, the fraction of indices $i \in \{1, \dots, N\}$ for which $I(W_N^{(i)}) \in (1-\delta, 1]$ goes to $I(W)$ and the fraction for which $I(W_N^{(i)}) \in [0, \delta)$ goes to $1 - I(W)$.

*Theorem 2* [1]: For any B-DMC $\{W\}$ with $I(W) > 0$, fixed $R < I(W)$, and fixed $\beta < \frac{1}{2}$, there exists a sequence of sets $\mathcal{A}_N \subseteq \{1, \dots, N\}, N \in \{1, 2, \dots, 2^n, \dots\}$, such that $|\mathcal{A}_N| \geq NR$ and $Z(W_N^{(i)}) \leq 2^{-N^\beta}$ for all $i \in \mathcal{A}_N$ and $N$ sufficiently large.

We can now take advantage of the polarization effect to construct codes that achieve the symmetric channel capacity $I(W)$.

*Definition 1*: Given a B-DMC $\{W: \mathcal{X} \to \mathcal{Y}\}$, let $N = 2^n$ for some $n \geq 0$ and let $\mathcal{A} \subseteq \{1, \dots, N\}$ and $u_{\mathcal{A}^c} \in \mathcal{X}^{|\mathcal{A}^c|}$ be fixed, an $N$-length $G_N$-coset code $(f_{G_N}, \varphi_{SC})$ under successive cancellation (SC) decoding for $\{W\}$ with respect to $\mathcal{A}$ and $u_{\mathcal{A}^c}$ is an $N$-length code for $\{W\}$ such that

i) $\mathcal{M} = \mathcal{M}_{f_{G_N}} = \mathcal{M}_{f_{G_N}}(u_{\mathcal{A}^c}) \triangleq \{m_1^N \in \mathcal{X}^N: m_{\mathcal{A}^c} = u_{\mathcal{A}^c}\};$

ii) $\widehat{\mathcal{M}} \triangleq \mathcal{M};$

iii) $f_{G_N}(m_1^N) \triangleq m_1^N G_N \in \mathcal{X}^N$ for all $m_1^N \in \mathcal{M}$ where $G_N$ is the generator matrix of order $N$;

iv) $\varphi_{SC}(y_1^N) \triangleq \widehat{m}_1^N \in \widehat{\mathcal{M}}$ for all $y_1^N \in \mathcal{Y}^N$ where the vector $\widehat{m}_1^N$ is generated by computing

$$\widehat{m}_i \triangleq \begin{cases} u_i, & \text{if } i \in \mathcal{A}^c \\ h_i(y_1^N, \widehat{m}_1^{i-1}), & \text{if } i \in \mathcal{A} \end{cases}$$

in the order $i$ from 1 to $N$ with decision functions $h_i: \mathcal{Y}^N \times \mathcal{X}^{i-1} \to \mathcal{X}, i \in \mathcal{A}$ defined as

$$h_i(y_1^N, \widehat{m}_1^{i-1}) \triangleq \begin{cases} 0, \text{if } \frac{W_N^{(i)}(y_1^N, \widehat{m}_1^{i-1}|0)}{W_N^{(i)}(y_1^N, \widehat{m}_1^{i-1}|1)} \geq 1 \\ 1, \text{otherwise} \end{cases}$$

for all $y_1^N \in \mathcal{Y}^N, \widehat{m}_1^{i-1} \in \mathcal{X}^{i-1}$.

We refer to $\mathcal{A}$ as the information set and to $u_{\mathcal{A}^c}$ as frozen bits or vector.

The rate of this code is $|\mathcal{A}|/N$. We will use the notation $\bar{e}(N, \mathcal{A}, u_{\mathcal{A}^c})$ to denote the average probability of error of this code, i.e.,

$$\bar{e}(N, \mathcal{A}, u_{\mathcal{A}^c}) \triangleq \bar{e}(W^N, f_{G_N}, \varphi_{SC}).$$

The average of $\bar{e}(N, \mathcal{A}, u_{\mathcal{A}^c})$ over all choices for $u_{\mathcal{A}^c}$ will be denoted by $\bar{e}(N, \mathcal{A})$, i.e.,

$$\bar{e}(N, \mathcal{A}) \triangleq \frac{1}{2^{|\mathcal{A}^c|}} \sum_{u_{\mathcal{A}^c} \in \mathcal{X}^{|\mathcal{A}^c|}} \bar{e}(N, \mathcal{A}, u_{\mathcal{A}^c}).$$

Then we have the bound [1]

$$\bar{e}(N, \mathcal{A}) \leq \sum_{i \in \mathcal{A}} Z(W_N^{(i)}).$$

This result suggests choosing $\mathcal{A}$ so as to minimize the right-hand side of above bound. This idea leads to the definition of polar codes.

*Definition 2*: An information set $\mathcal{A}$ is called a polar set for $\{W\}$ if $Z(W_N^{(i)}) \leq Z(W_N^{(j)})$ for all $i \in \mathcal{A}$, $j \in \mathcal{A}^c$. An $N$-length $G_N$-coset code under SC decoding for $\{W\}$ with respect to $\mathcal{A}$ and $u_{\mathcal{A}^c}$ is called an $N$-length polar code under SC decoding for $\{W\}$ with respect to $\mathcal{A}$ and $u_{\mathcal{A}^c}$ if $\mathcal{A}$ is a polar set for $\{W\}$.

For a number $R \geq 0$, let $\bar{e}(N, R)$ be defined as $\bar{e}(N, \mathcal{A})$ where $\mathcal{A}$ is a polar set with $|\mathcal{A}| = \lfloor NR \rfloor$. Thus, $\bar{e}(N, R)$ is the average probability of error of an $N$-length polar code for $\{W\}$ with rate $R$, averaged over all choices for the frozen bits $u_{\mathcal{A}^c}$. Then we have the following coding theorem.

*Theorem 3* [1]: For any given B-DMC $\{W\}$, fixed $R < I(W)$, and fixed $\beta < \frac{1}{2}$, average error probability for $N$-length polar code under SC decoding satisfies

$$\bar{e}(N, R) \leq 2^{-N^\beta}$$

for $N$ sufficiently large.

This theorem shows that the polar codes under SC decoding can achieve the symmetric capacity $I(W)$ of any given B-DMC $\{W\}$.

The symmetric capacity $I(W)$ equals the Shannon capacity when $W$ is a symmetric channel, i.e., a channel for which there exists a permutation $\pi_1$ on the output alphabet $\mathcal{Y}$ such that $\pi_1^{-1} = \pi_1$ and $W(y|1) = W(\pi_1(y)|0)$ for all $y \in \mathcal{Y}$. Let $\pi_0$ be the identity permutation on $\mathcal{Y}$. Clearly, the permutations $(\pi_0, \pi_1)$ form an Abelian group under function composition. For a compact notation, we will write $x \cdot y$ to denote $\pi_x(y)$ for $x \in \mathcal{X}, y \in \mathcal{Y}$. For $x_1^N \in \mathcal{X}^N, y_1^N \in \mathcal{Y}^N$, let

$$x_1^N \cdot y_1^N \triangleq (x_1 \cdot y_1, \ldots, x_N \cdot y_N).$$

This associates to each element of $\mathcal{X}^N$ a permutation on $\mathcal{Y}^N$.

If a B-DMC $\{W\}$ is symmetric, then the channels $W_N$ and $W_N^{(i)}$ are also symmetric in the sense that [1]

$$W_N(y_1^N|m_1^N) = W_N(a_1^N G_N \cdot y_1^N|m_1^N \oplus a_1^N), \quad (1)$$

$$W_N^{(i)}(y_1^N, m_1^{i-1}|m_i) = W_N^{(i)}(a_1^N G_N \cdot y_1^N, m_1^{i-1} \oplus a_1^{i-1}|m_i \oplus a_i) \quad (2)$$

for all $m_1^N, a_1^N \in \mathcal{X}^N, y_1^N \in \mathcal{Y}^N, N = 2^n, n \geq 0, 1 \leq i \leq N$.

For symmetric channels, we have the following stronger version of Theorem 3.

*Theorem 4* [1]: For any symmetric B-DMC $\{W\}$, fixed $R < I(W)$, and fixed $\beta < \frac{1}{2}$, consider any sequence of polar codes $(f_{G_N}, \varphi_{SC})$ under SC decoding with $N$ increasing to infinity, $|\mathcal{A}| = \lfloor NR \rfloor$, and $u_{\mathcal{A}^c}$ fixed arbitrarily. The average error probability satisfies

$$\bar{e}(N, \mathcal{A}, u_{\mathcal{A}^c}) \leq 2^{-N^\beta}$$

for $N$ sufficiently large.

Note that for symmetric channels $I(W)$ equals the Shannon capacity. This theorem shows that, for symmetric channels, any choice for $u_{\mathcal{A}^c}$ is as good as any other.

Although the polar codes under SC decoding achieve capacity asymptotically, empirical studies indicate that for short and moderate block lengths, they do not perform as well as turbo codes or low density parity check (LDPC) codes. Thus, either the encoder or the decoder is weak. In fact, the two causes are complementary and both contribute to the problem. Tal and Vardy proposed two improvements respectively to the encoder and the decoder in [2].

*Improvement 1* [2]: A successive cancellation list (SCL) decoder is proposed which is a generalization of the SC decoder of [1]. The SCL decoder is governed by a single integer parameter $L$, which denotes the list size. As in the SC decoder, the input bits are decoded successively one-by-one. However, in the SCL decoder, $L$ decoding paths are considered concurrently at each decoding stage. Specifically, the SCL decoder doubles the number of decoding paths for each information bit, and then uses a pruning procedure to discard all but the $L$ most likely paths. At the end of the decoding process, the most likely among the $L$ decoding paths is selected as the decoder output. Simulation results show that for a wide range of polar codes of various lengths, SCL decoding effectively bridges the performance gap between SC decoding and maximum-likelihood (ML) decoding.

*Improvement 2* [2]: It has been observed in simulations that even under ML decoding, the performance of polar codes falls short in comparison to LDPC and turbo codes of comparable length, which demonstrates the weakness of the encoder. It has been noticed in simulations that, under SCL decoding, with high probability, the transmitted message is on the list, but it is not the most likely message on the list. It is therefore not selected as the decoder output. This means that performance could be further improved if we had a genie aided decoder capable of identifying the transmitted message whenever it is on the list. But such a genie can be easily implemented, for example using cyclic redundancy check (CRC) precoding. It has been confirmed in numerous simulations that the performance of polar codes under SCL decoding with CRC is comparable to state-of-the-art turbo and LDPC codes.

Although the performance of the codes in [2] is encouraging in practice, it is not known whether they are good in theory as well. Some thought shows that the likelihood probability of the decoding output of the SC decoder is not necessarily less than or equal to that of the SCL decoder. So the problem is not trivial. In this paper we address this problem. We will show that the codes in both Improvement 1 and Improvement 2 can achieve the symmetric capacity $I(W)$ of any given B-DMC $\{W\}$ asymptotically.

The rest of the paper is organized as follows. In Section II, we formalize the codes in Improvement 1, i.e., the polar codes under SCL decoding, with our notation and prove that they achieve the symmetric capacity $I(W)$ of any given B-DMC $\{W\}$ asymptotically, which is our one main result. In Section III, we consider the polar codes under SCL decoding for symmetric B-DMCs and prove a stronger result. In Section IV, we formalize the codes in Improvement 2. In fact, we will propose a family of more general codes, i.e., precoded polar codes under SCL decoding and prove that they achieve the symmetric capacity $I(W)$ of any given B-DMC $\{W\}$ asymptotically under some conditions, which is our another main result. In Section V, we give some extensions about our results and draw a conclusion.

## II. SUCCESSIVE CANCELLATION LIST DECODING

In this section, we recast the SCL decoder using our notation, for future proof. Let $\mathbb{N}$ denote the positive integer set. For a finite set $\mathcal{A}$, let $\mathbb{S}_{\mathcal{A}}$ denote the collection of all nonempty subsets of $\mathcal{A}$.

*Definition 3*: Given any discrete channel $W: \mathcal{X} \to \mathcal{Y}$, a most-likely mapping $g_W: \mathbb{S}_{\mathcal{Y}} \times \mathbb{S}_{\mathcal{X}} \times \mathbb{N} \to \mathbb{S}_{\mathcal{Y} \times \mathcal{X}}$ is defined as

$$g_W(\mathcal{T}, \mathcal{S}, L) \triangleq \mathcal{Z}$$

for all $\mathcal{T} \in \mathbb{S}_{\mathcal{Y}}, \mathcal{S} \in \mathbb{S}_{\mathcal{X}}, L \in \mathbb{N}$, where $\mathcal{Z} \in \mathbb{S}_{\mathcal{Y} \times \mathcal{X}}$ satisfies that
 if $|\mathcal{T} \times \mathcal{S}| \leq L$, then $\mathcal{Z} = \mathcal{T} \times \mathcal{S}$;
 if $|\mathcal{T} \times \mathcal{S}| > L$, then $\mathcal{Z}$ is chosen as the $L$-element subset of $\mathcal{T} \times \mathcal{S}$ such that $W(t|s) \geq W(y|x)$ for all $(t, s) \in \mathcal{Z}$ and $(y, x) \in (\mathcal{T} \times \mathcal{S}) \backslash \mathcal{Z}$. In addition, since $\mathcal{X}$ and $\mathcal{Y}$ are finite, there exists a strict total order $\prec$ on the set $\mathcal{Y} \times \mathcal{X}$. Then for all $(t,s) \in \mathcal{Z}$ and $(y,x) \in (\mathcal{T} \times \mathcal{S}) \backslash \mathcal{Z}$ such that $W(t|s) = W(y|x)$, $(t,s) \prec (y,x)$ holds.

Thus, in word, what a most-likely mapping does is to select from $|\mathcal{T} \times \mathcal{S}|$ input-output pairs $L$ most likely ones, giving it the name. The purpose of introducing the strict total order relation to Definition 3 is just to guarantee that the mapping is well-defined.

*Definition 4*: Given a B-DMC $\{W: \mathcal{X} \to \mathcal{Y}\}$, let $N = 2^n$ for some $n \geq 0$ and $L \in \mathbb{N}$ be fixed, an $N$-length $G_N$-coset code $(f_{G_N}, \varphi_{L-\text{SCL}})$ under $L$-SCL decoding for $\{W\}$ with respect to $\mathcal{A}$ and $u_{\mathcal{A}^c}$ is an $N$-length code for $\{W\}$ where $\mathcal{M}, \widehat{\mathcal{M}}, f_{G_N}, \mathcal{A},$ and $u_{\mathcal{A}^c}$ are defined as in Definition 1 and the decoder $\varphi_{L-\text{SCL}}$ consists of a successive-cancellation-list generating phase and a local-maximum-likelihood selecting phase.

1) Successive-Cancellation-List Generating (SCLG): For any channel output $y_1^N \in \mathcal{Y}^N$, this phase produces a vector set $\widehat{\mathcal{M}}_1^N(u_{\mathcal{A}^c}, y_1^N) \subseteq \widehat{\mathcal{M}}$ in a recursive manner by computing

$\widehat{\mathcal{M}}_1^i(u_{\mathcal{A}_i^c}, y_1^N)$

$\triangleq \begin{cases} \{\widehat{m}_1^i : \widehat{m}_1^{i-1} \in \widehat{\mathcal{M}}_1^{i-1}(u_{\mathcal{A}_{i-1}^c}, y_1^N), \widehat{m}_i = u_i\} & \text{if } i \in \mathcal{A}^c \\ \{\widehat{m}_1^i : (y_1^N, \widehat{m}_1^i) \in g_{W_N^{(i)}}(\{y_1^N\} \times \widehat{\mathcal{M}}_1^{i-1}(u_{\mathcal{A}_{i-1}^c}, y_1^N), \mathcal{X}, L)\} & \text{if } i \in \mathcal{A} \end{cases}$

in the order $i$ from 1 to $N$ where $\mathcal{A}_i^c \triangleq \mathcal{A}^c \cap \{1, \dots, i\}$. Note that $\mathcal{A}_N^c = \mathcal{A}^c$.

2) Local-Maximum-Likelihood Selecting (LMLS): Having generated the vector set $\widehat{\mathcal{M}}_1^N(u_{\mathcal{A}^c}, y_1^N)$ from the channel output $y_1^N$, this phase of the decoder selects a most likely vector from $\widehat{\mathcal{M}}_1^N(u_{\mathcal{A}^c}, y_1^N)$ by computing

$\{(y_1^N, \widehat{m}_1^N)\} \triangleq g_{W_N}(\{y_1^N\}, \widehat{\mathcal{M}}_1^N(u_{\mathcal{A}^c}, y_1^N), 1).$

Then $\varphi_{L-\text{SCL}}(y_1^N) \triangleq \widehat{m}_1^N \in \widehat{\mathcal{M}}$ for all $y_1^N \in \mathcal{Y}^N$.

*Remark*: One can verify readily that the decoder $\varphi_{L-\text{SCL}}$ defined above is indeed the SCL decoder with list size $L$ described in [2]. Specifically, the SCLG phase formalizes the process that doubles the number of decoding paths for each information bit, and then uses a pruning g procedure to discard all but the $L$ most likely paths. The LMLS phase formalizes the last step of the decoding process, i.e., the most likely among the $L$ decoding paths is selected as the decoder output.

Given an $N$-length code $(f_{G_N}, \varphi_{L-\text{SCL}})$ for $\{W\}$ with respect to an information set $\mathcal{A}$ and some frozen bits $u_{\mathcal{A}^c}$, we will use the notation $\bar{e}(N, L, \mathcal{A}, u_{\mathcal{A}^c})$ to denote the average probability of error of the code, i.e.,

$\bar{e}(N, L, \mathcal{A}, u_{\mathcal{A}^c}) \triangleq \bar{e}(W^N, f_{G_N}, \varphi_{L-\text{SCL}}).$

The average of $\bar{e}(N, L, \mathcal{A}, u_{\mathcal{A}^c})$ over all choices for $u_{\mathcal{A}^c}$ will be denoted by $\bar{e}(N, L, \mathcal{A})$, i.e.,

$\bar{e}(N, L, \mathcal{A}) \triangleq \frac{1}{2^{|\mathcal{A}^c|}} \sum_{u_{\mathcal{A}^c} \in \mathcal{X}^{|\mathcal{A}^c|}} \bar{e}(N, L, \mathcal{A}, u_{\mathcal{A}^c}).$

*Definition 5*: An $N$-length $G_N$-coset code under $L$-SCL decoding for $\{W\}$ with respect to $\mathcal{A}$ and $u_{\mathcal{A}^c}$ is called an $N$-length polar code under $L$-SCL decoding for $\{W\}$ with respect to $\mathcal{A}$ and $u_{\mathcal{A}^c}$ if $\mathcal{A}$ is a polar set for $\{W\}$.

For a number $R \geq 0$, let $\bar{e}(N, R, L)$ be defined as $\bar{e}(N, L, \mathcal{A})$ where $\mathcal{A}$ is a polar set with $|\mathcal{A}| = \lfloor NR \rfloor$. Thus, $\bar{e}(N, R, L)$ is the average probability of error of an $N$-length polar code under $L$-SCL decoding for $\{W\}$ with rate $R$, averaged over all choices for the frozen bits $u_{\mathcal{A}^c}$.

Now our first main result is to show that polar codes under $L$-SCL decoding can achieve the symmetric capacity $I(W)$ of any given B-DMC $\{W\}$.

*Theorem 5*: For any given B-DMC $\{W\}$, fixed $R < I(W)$, $L \geq 1$, and $\beta < \frac{1}{2}$, average error probability for $N$-length polar code under $L$-SCL decoding satisfies

$$\bar{e}(N, R, L) \leq 2^{-N^\beta}$$

for $N$ sufficiently large.

Consider an $N$-length $G_N$-coset code $(f_{G_N}, \varphi_{L-\text{SCL}})$ under $L$-SCL decoding for $\{W: \mathcal{X} \to \mathcal{Y}\}$ with respect to $\mathcal{A}$ and $u_{\mathcal{A}^c}$. Let a message $m_1^N$ be encoded into a codeword $x_1^N$, let $x_1^N$ be sent over the channel $W^N$, let a channel output $y_1^N$ be received, and let $\widehat{\mathcal{M}}_1^N(u_{\mathcal{A}^c}, y_1^N)$ be the output of the SCLG phase of the decoder, see Definition 4. We will use the notation $P_e(N, L, \mathcal{A}, u_{\mathcal{A}^c})$ to denote the average probability of $m_1^N \notin \widehat{\mathcal{M}}_1^N(u_{\mathcal{A}^c}, y_1^N)$, assuming that each message $m_1^N$ is sent with probability $2^{-|\mathcal{A}|}$. More precisely,

$P_e(N, L, \mathcal{A}, u_{\mathcal{A}^c})$

$\triangleq \sum_{m_1^N \in \mathcal{M}_{f_{G_N}}(u_{\mathcal{A}^c})} \frac{1}{2^{|\mathcal{A}|}} \sum_{y_1^N \in \mathcal{Y}^N : m_1^N \notin \widehat{\mathcal{M}}_1^N(u_{\mathcal{A}^c}, y_1^N)} W_N(y_1^N | m_1^N). \quad (3)$

The average of $P_e(N, L, \mathcal{A}, u_{\mathcal{A}^c})$ over all choices for $u_{\mathcal{A}^c}$ will be denoted by $P_e(N, L, \mathcal{A})$, i.e.,

$P_e(N, L, \mathcal{A}) \triangleq \sum_{u_{\mathcal{A}^c} \in \mathcal{X}^{|\mathcal{A}^c|}} \frac{1}{2^{|\mathcal{A}^c|}} P_e(N, L, \mathcal{A}, u_{\mathcal{A}^c}). \quad (4)$

First, we will prove the following bound.

*Proposition 1*: For any B-DMC $\{W\}$ and any choice of the parameters $N, L, \mathcal{A}$

$$P_e(N, L, \mathcal{A}) \leq \sum_{i \in \mathcal{A}} Z(W_N^{(i)}).$$

*Proof.* From (3) and (4), we know

$P_e(N, L, \mathcal{A}) = \sum_{u_{\mathcal{A}^c} \in \mathcal{X}^{|\mathcal{A}^c|}} \frac{1}{2^{|\mathcal{A}^c|}} P_e(N, L, \mathcal{A}, u_{\mathcal{A}^c})$

$= \sum_{m_1^N \in \mathcal{X}^N} \sum_{y_1^N \in \mathcal{Y}^N : m_1^N \notin \widehat{\mathcal{M}}_1^N(m_{\mathcal{A}^c}, y_1^N)} \frac{1}{2^N} W_N(y_1^N | m_1^N)$

$= \sum_{y_1^N \in \mathcal{Y}^N} \sum_{m_1^N \in \mathcal{X}^N : m_1^N \notin \widehat{\mathcal{M}}_1^N(m_{\mathcal{A}^c}, y_1^N)} \frac{1}{2^N} W_N(y_1^N | m_1^N)$

$$= \sum_{y_1^N \in \mathcal{Y}^N} \sum_{m_1^N \in \mathcal{E}(y_1^N)} \frac{1}{2^N} W_N(y_1^N | m_1^N)$$

where the set $\mathcal{E}(y_1^N)$ is defined as

$$\mathcal{E}(y_1^N) \triangleq \{m_1^N \in \mathcal{X}^N : m_1^N \notin \widehat{\mathcal{M}}_1^N(m_{\mathcal{A}^c}, y_1^N)\}$$

for all $y_1^N \in \mathcal{Y}^N$.

For every $i \in \mathcal{A}$, define

$$\mathcal{E}_i(y_1^N) \triangleq \{m_1^N \in \mathcal{X}^N : m_1^{i-1} \in \widehat{\mathcal{M}}_1^{i-1}(m_{\mathcal{A}_{i-1}^c}, y_1^N),$$
$$m_1^i \notin \widehat{\mathcal{M}}_1^i(m_{\mathcal{A}_i^c}, y_1^N)\}. \quad (5)$$

Then it is clear that

$$P_e(N, L, \mathcal{A}) = \sum_{y_1^N \in \mathcal{Y}^N} \sum_{m_1^N \in \mathcal{E}(y_1^N)} \frac{1}{2^N} W_N(y_1^N | m_1^N)$$
$$= \sum_{y_1^N \in \mathcal{Y}^N} \sum_{i \in \mathcal{A}} \sum_{m_1^N \in \mathcal{E}_i(y_1^N)} \frac{1}{2^N} W_N(y_1^N | m_1^N).$$

For any $i \in \mathcal{A}$, frozen bits $u_{\mathcal{A}^c}$, and $y_1^N \in \mathcal{Y}^N$, we can partition the set $\widehat{\mathcal{M}}_1^{i-1}(u_{\mathcal{A}_{i-1}^c}, y_1^N)$ into three subsets, i.e.,

$$\widehat{\mathcal{M}}_{i-1}^{(0)}\left(u_{\mathcal{A}_{i-1}^c}, y_1^N\right) \triangleq \{\widehat{m}_1^{i-1} \in \widehat{\mathcal{M}}_1^{i-1}\left(u_{\mathcal{A}_{i-1}^c}, y_1^N\right) :$$
$$(\widehat{m}_1^{i-1}, 0) \notin \widehat{\mathcal{M}}_1^i\left(u_{\mathcal{A}_i^c}, y_1^N\right), (\widehat{m}_1^{i-1}, 1) \notin \widehat{\mathcal{M}}_1^i\left(u_{\mathcal{A}_i^c}, y_1^N\right)\}, \quad (6)$$

$$\widehat{\mathcal{M}}_{i-1}^{(1)}(u_{\mathcal{A}_{i-1}^c}, y_1^N) \triangleq \{\widehat{m}_1^{i-1} \in \widehat{\mathcal{M}}_1^{i-1}\left(u_{\mathcal{A}_{i-1}^c}, y_1^N\right) :$$
$$(\widehat{m}_1^{i-1}, b) \notin \widehat{\mathcal{M}}_1^i\left(u_{\mathcal{A}_i^c}, y_1^N\right), (\widehat{m}_1^{i-1}, b\oplus 1) \in \widehat{\mathcal{M}}_1^i\left(u_{\mathcal{A}_i^c}, y_1^N\right)\}, \quad (7)$$

and

$$\widehat{\mathcal{M}}_{i-1}^{(2)}(u_{\mathcal{A}_{i-1}^c}, y_1^N) \triangleq \{\widehat{m}_1^{i-1} \in \widehat{\mathcal{M}}_1^{i-1}\left(u_{\mathcal{A}_{i-1}^c}, y_1^N\right) :$$
$$(\widehat{m}_1^{i-1}, 0) \in \widehat{\mathcal{M}}_1^i\left(u_{\mathcal{A}_i^c}, y_1^N\right), (\widehat{m}_1^{i-1}, 1) \in \widehat{\mathcal{M}}_1^i\left(u_{\mathcal{A}_i^c}, y_1^N\right)\}. \quad (8)$$

It is obvious that they are disjoint with each other and

$$\widehat{\mathcal{M}}_1^{i-1}(u_{\mathcal{A}_{i-1}^c}, y_1^N)$$
$$= \widehat{\mathcal{M}}_{i-1}^{(0)}(u_{\mathcal{A}_{i-1}^c}, y_1^N) \cup \widehat{\mathcal{M}}_{i-1}^{(1)}(u_{\mathcal{A}_{i-1}^c}, y_1^N) \cup \widehat{\mathcal{M}}_{i-1}^{(2)}(u_{\mathcal{A}_{i-1}^c}, y_1^N).$$

Thus, we have

$$\left|\widehat{\mathcal{M}}_1^{i-1}\left(u_{\mathcal{A}_{i-1}^c}, y_1^N\right)\right| = \left|\widehat{\mathcal{M}}_{i-1}^{(0)}\left(u_{\mathcal{A}_{i-1}^c}, y_1^N\right)\right|$$
$$+ \left|\widehat{\mathcal{M}}_{i-1}^{(1)}\left(u_{\mathcal{A}_{i-1}^c}, y_1^N\right)\right| + \left|\widehat{\mathcal{M}}_{i-1}^{(2)}\left(u_{\mathcal{A}_{i-1}^c}, y_1^N\right)\right|. \quad (9)$$

In addition, from (6), (7), and (8), we also know

$$\left|\widehat{\mathcal{M}}_1^i\left(u_{\mathcal{A}_i^c}, y_1^N\right)\right|$$
$$= \left|\widehat{\mathcal{M}}_{i-1}^{(1)}\left(u_{\mathcal{A}_{i-1}^c}, y_1^N\right)\right| + 2\left|\widehat{\mathcal{M}}_{i-1}^{(2)}\left(u_{\mathcal{A}_{i-1}^c}, y_1^N\right)\right|. \quad (10)$$

From Definition 4 and Definition 3, it holds that

$$\left|\widehat{\mathcal{M}}_1^{i-1}\left(u_{\mathcal{A}_{i-1}^c}, y_1^N\right)\right| \leq \left|\widehat{\mathcal{M}}_1^i\left(u_{\mathcal{A}_i^c}, y_1^N\right)\right|.$$

Therefore, by (9) and (10), we have

$$\left|\widehat{\mathcal{M}}_{i-1}^{(0)}(u_{\mathcal{A}_{i-1}^c}, y_1^N)\right| \leq \left|\widehat{\mathcal{M}}_{i-1}^{(2)}\left(u_{\mathcal{A}_{i-1}^c}, y_1^N\right)\right|. \quad (11)$$

From (7), Definition 4, and Definition 3, it holds that, for all $\widehat{m}_1^{i-1} \in \widehat{\mathcal{M}}_{i-1}^{(1)}(u_{\mathcal{A}_{i-1}^c}, y_1^N)$,

$$W_N^{(i)}(y_1^N, \widehat{m}_1^{i-1} | b) \leq W_N^{(i)}(y_1^N, \widehat{m}_1^{i-1} | b \oplus 1). \quad (12)$$

From (6), (8), Definition 4, and Definition 3, it holds that

$$W_N^{(i)}(y_1^N, m_1^{i-1} | b') \leq W_N^{(i)}(y_1^N, \widetilde{m}_1^{i-1} | \widetilde{b}') \quad (13)$$

for all $m_1^{i-1} \in \widehat{\mathcal{M}}_{i-1}^{(0)}(u_{\mathcal{A}_{i-1}^c}, y_1^N)$, $\widetilde{m}_1^{i-1} \in \widehat{\mathcal{M}}_{i-1}^{(2)}\left(u_{\mathcal{A}_{i-1}^c}, y_1^N\right)$, and all $b', \widetilde{b}' \in \mathcal{X}$. Thus,

$$P_e(N, L, \mathcal{A})$$
$$= \sum_{y_1^N \in \mathcal{Y}^N} \sum_{i \in \mathcal{A}} \sum_{m_1^N \in \mathcal{E}_i(y_1^N)} \frac{1}{2^N} W_N(y_1^N | m_1^N)$$
$$= \sum_{y_1^N \in \mathcal{Y}^N} \sum_{i \in \mathcal{A}} \Big( \sum_{\substack{m_1^{i-1} \in \mathcal{X}^{i-1}: \\ m_1^{i-1} \in \widehat{\mathcal{M}}_{i-1}^{(1)}(m_{\mathcal{A}_{i-1}^c}, y_1^N)}} \frac{1}{2} W_N^{(i)}(y_1^N, m_1^{i-1} | b)$$
$$+ \sum_{\substack{m_1^{i-1} \in \mathcal{X}^{i-1}: \\ m_1^{i-1} \in \widehat{\mathcal{M}}_{i-1}^{(0)}(m_{\mathcal{A}_{i-1}^c}, y_1^N), \\ b' \in \mathcal{X}}} \frac{1}{2} W_N^{(i)}(y_1^N, m_1^{i-1} | b')\Big)$$
$$\leq \sum_{y_1^N \in \mathcal{Y}^N} \sum_{i \in \mathcal{A}} \Big( \sum_{\substack{m_1^{i-1} \in \mathcal{X}^{i-1}: \\ m_1^{i-1} \in \widehat{\mathcal{M}}_{i-1}^{(1)}(m_{\mathcal{A}_{i-1}^c}, y_1^N)}} \frac{1}{2} \sqrt{W_N^{(i)}(y_1^N, m_1^{i-1} | 0)} \cdot \sqrt{W_N^{(i)}(y_1^N, m_1^{i-1} | 1)}$$
$$+ \sum_{\substack{m_1^{i-1} \in \mathcal{X}^{i-1}: \\ m_1^{i-1} \in \widehat{\mathcal{M}}_{i-1}^{(2)}(m_{\mathcal{A}_{i-1}^c}, y_1^N)}} \sqrt{W_N^{(i)}(y_1^N, m_1^{i-1} | 0) W_N^{(i)}(y_1^N, m_1^{i-1} | 1)}\Big)$$
$$\leq \sum_{i \in \mathcal{A}} \sum_{y_1^N \in \mathcal{Y}^N, m_1^{i-1} \in \mathcal{X}^{i-1}} \sqrt{W_N^{(i)}(y_1^N, m_1^{i-1} | 0) W_N^{(i)}(y_1^N, m_1^{i-1} | 1)}$$
$$= \sum_{i \in \mathcal{A}} Z(W_N^{(i)})$$

where the second equality is due to (5), (6), (7), and (8), the first inequality follows from (11), (12), and (13), and the second inequality is obtained by noting that

$$\{m_1^{i-1} \in \mathcal{X}^{i-1} : m_1^{i-1} \in \widehat{\mathcal{M}}_{i-1}^{(1)}\left(m_{\mathcal{A}_{i-1}^c}, y_1^N\right)\}$$
$$\cap \{m_1^{i-1} \in \mathcal{X}^{i-1} : m_1^{i-1} \in \widehat{\mathcal{M}}_{i-1}^{(2)}\left(m_{\mathcal{A}_{i-1}^c}, y_1^N\right)\} = \emptyset$$

and

$$\{m_1^{i-1} \in \mathcal{X}^{i-1} : m_1^{i-1} \in \widehat{\mathcal{M}}_{i-1}^{(1)}\left(m_{\mathcal{A}_{i-1}^c}, y_1^N\right)\}$$
$$\cup \{m_1^{i-1} \in \mathcal{X}^{i-1} : m_1^{i-1} \in \widehat{\mathcal{M}}_{i-1}^{(2)}\left(m_{\mathcal{A}_{i-1}^c}, y_1^N\right)\} \subseteq \mathcal{X}^{i-1}.$$

This completes the proof of Proposition 1.

Second, we will prove the following bound.

*Proposition 2*: For any B-DMC $\{W\}$ and any choice of the parameters $N, L, \mathcal{A}$

$$\bar{e}(N, L, \mathcal{A}) \leq 2 \sum_{i \in \mathcal{A}} Z(W_N^{(i)}).$$

*Proof*: Given a B-DMC $\{W: \mathcal{X} \to \mathcal{Y}\}$, let $N = 2^n$ for some $n \geq 0$ and $(f_{G_N}, \varphi_{L\text{-SCL}})$ be an $N$-length $G_N$-coset code under $L$-SCL decoding for $\{W\}$ with respect to $\mathcal{A}$ and $u_{\mathcal{A}^c}$. Define

$$\mathcal{Y}_L(m_1^N) \triangleq \{y_1^N \in \mathcal{Y}^N : m_1^N \in \widehat{\mathcal{M}}_1^N(u_{\mathcal{A}^c}, y_1^N)\} \quad (14)$$

for all messages $m_1^N \in \mathcal{M}_{f_{G_N}}(u_{\mathcal{A}^c})$.

We note that among all $L \geq 1$, $L = 1$ is special in the sense that $\mathcal{Y}_1(m_1^N) = \varphi_{1\text{-SCL}}^{-1}(m_1^N)$. Thus, $\mathcal{Y}^N$ is a disjoint union of the sets $\mathcal{Y}_1(m_1^N)$ for all messages $m_1^N \in \mathcal{M}_{f_{G_N}}(u_{\mathcal{A}^c})$ and by Proposition 1

$$1 - \bar{e}(N, 1, \mathcal{A})$$
$$= \sum_{u_{\mathcal{A}^c} \in \mathcal{X}^{|\mathcal{A}^c|}} \sum_{m_1^N \in \mathcal{M}_{f_{G_N}}(u_{\mathcal{A}^c})} \frac{1}{2^N} W_N(\varphi_{1\text{-SCL}}^{-1}(m_1^N)|m_1^N)$$
$$= \sum_{u_{\mathcal{A}^c} \in \mathcal{X}^{|\mathcal{A}^c|}} \sum_{m_1^N \in \mathcal{M}_{f_{G_N}}(u_{\mathcal{A}^c})} \frac{1}{2^N} W_N(\mathcal{Y}_1(m_1^N)|m_1^N)$$
$$= 1 - P_e(N, 1, \mathcal{A})$$
$$\geq 1 - \sum_{i \in \mathcal{A}} Z(W_N^{(i)}).$$

Thus we get $\bar{e}(N, 1, \mathcal{A}) \leq \sum_{i \in \mathcal{A}} Z(W_N^{(i)})$ as desired.

But when $L > 1$, it is obvious that $\mathcal{Y}_L(m_1^N) \supseteq \varphi_{L\text{-SCL}}^{-1}(m_1^N)$, the sets $\mathcal{Y}_L(m_1^N)$ cover $\mathcal{Y}^N$ but are not necessarily pairwise disjoint, and

$$1 - \bar{e}(N, L, \mathcal{A})$$
$$= \sum_{u_{\mathcal{A}^c} \in \mathcal{X}^{|\mathcal{A}^c|}} \sum_{m_1^N \in \mathcal{M}_{f_{G_N}}(u_{\mathcal{A}^c})} \frac{1}{2^N} W_N(\varphi_{L\text{-SCL}}^{-1}(m_1^N)|m_1^N)$$
$$\leq \sum_{u_{\mathcal{A}^c} \in \mathcal{X}^{|\mathcal{A}^c|}} \sum_{m_1^N \in \mathcal{M}_{f_{G_N}}(u_{\mathcal{A}^c})} \frac{1}{2^N} W_N(\mathcal{Y}_L(m_1^N)|m_1^N)$$
$$= 1 - P_e(N, L, \mathcal{A})$$
$$\geq 1 - \sum_{i \in \mathcal{A}} Z(W_N^{(i)})$$

from which we cannot obtain what we want.

For any fixed $L \geq 1$, define

$$\mathcal{Y}(m_1^N) \triangleq \mathcal{Y}_1(m_1^N) \cap \mathcal{Y}_L(m_1^N)$$

for all messages $m_1^N \in \mathcal{M}_{f_{G_N}}(u_{\mathcal{A}^c})$. Then, by Proposition 1,

$$1 - \sum_{u_{\mathcal{A}^c} \in \mathcal{X}^{|\mathcal{A}^c|}} \sum_{m_1^N \in \mathcal{M}_{f_{G_N}}(u_{\mathcal{A}^c})} \frac{1}{2^N} W_N(\mathcal{Y}(m_1^N)|m_1^N)$$

$$= \sum_{u_{\mathcal{A}^c} \in \mathcal{X}^{|\mathcal{A}^c|}} \sum_{m_1^N \in \mathcal{M}_{f_{G_N}}(u_{\mathcal{A}^c})} \frac{1}{2^N} W_N(\mathcal{Y}^c(m_1^N)|m_1^N)$$

$$= \sum_{u_{\mathcal{A}^c} \in \mathcal{X}^{|\mathcal{A}^c|}} \sum_{m_1^N \in \mathcal{M}_{f_{G_N}}(u_{\mathcal{A}^c})} \frac{1}{2^N} W_N(\mathcal{Y}_1^c(m_1^N) \cup \mathcal{Y}_L^c(m_1^N)|m_1^N)$$

$$\leq \sum_{u_{\mathcal{A}^c} \in \mathcal{X}^{|\mathcal{A}^c|}} \sum_{m_1^N \in \mathcal{M}_{f_{G_N}}(u_{\mathcal{A}^c})} (\frac{1}{2^N} W_N(\mathcal{Y}_1^c(m_1^N)|m_1^N)$$
$$+ \frac{1}{2^N} W_N(\mathcal{Y}_L^c(m_1^N)|m_1^N))$$

$$= P_e(N, 1, \mathcal{A}) + P_e(N, L, \mathcal{A})$$

$$\leq 2 \sum_{i \in \mathcal{A}} Z(W_N^{(i)}). \quad (15)$$

As $\mathcal{Y}(m_1^N) \subseteq \mathcal{Y}_1(m_1^N)$, the sets $\mathcal{Y}(m_1^N)$ are pairwise disjoint and the union of them is a subset of $\mathcal{Y}^N$. On the other hand, since $\mathcal{Y}(m_1^N) \subseteq \mathcal{Y}_L(m_1^N)$, by (14) and Definition 4, for all $y_1^N \in \mathcal{Y}(m_1^N)$, we have $y_1^N \in \mathcal{Y}_L(m_1^N)$ which implies

$$W_N(y_1^N|m_1^N) \leq W_N(y_1^N|\varphi_{L\text{-SCL}}(y_1^N)). \quad (16)$$

Hence, it follows that

$$1 - \bar{e}(N, L, \mathcal{A})$$
$$= \sum_{u_{\mathcal{A}^c} \in \mathcal{X}^{|\mathcal{A}^c|}} \sum_{y_1^N \in \mathcal{Y}^N} \frac{1}{2^N} W_N(y_1^N|\varphi_{L\text{-SCL}}(y_1^N))$$
$$\geq \sum_{u_{\mathcal{A}^c} \in \mathcal{X}^{|\mathcal{A}^c|}} \sum_{m_1^N \in \mathcal{M}_{f_{G_N}}(u_{\mathcal{A}^c})} \sum_{y_1^N \in \mathcal{Y}(m_1^N)} \frac{1}{2^N} W_N(y_1^N|\varphi_{L\text{-SCL}}(y_1^N))$$
$$\geq \sum_{u_{\mathcal{A}^c} \in \mathcal{X}^{|\mathcal{A}^c|}} \sum_{m_1^N \in \mathcal{M}_{f_{G_N}}(u_{\mathcal{A}^c})} \sum_{y_1^N \in \mathcal{Y}(m_1^N)} \frac{1}{2^N} W_N(y_1^N|m_1^N)$$
$$= \sum_{u_{\mathcal{A}^c} \in \mathcal{X}^{|\mathcal{A}^c|}} \sum_{m_1^N \in \mathcal{M}_{f_{G_N}}(u_{\mathcal{A}^c})} \frac{1}{2^N} W_N(\mathcal{Y}(m_1^N)|m_1^N)$$
$$\geq 1 - 2 \sum_{i \in \mathcal{A}} Z(W_N^{(i)})$$

where the first inequality is due to the fact that the sets $\mathcal{Y}(m_1^N)$ are pairwise disjoint and the union of them is a subset of $\mathcal{Y}^N$, the second inequality is due to (16), and the last inequality is due to (15). This completes the proof of Proposition 2.

*Proof of Theorem 5*: By Theorem 2, for any given rate $R < I(W)$ and fixed $\beta < \frac{1}{2}$, there exists a sequence of information sets $\mathcal{A}_N$ with size $|\mathcal{A}_N| \geq NR$ such that

$$\sum_{i \in \mathcal{A}_N} Z(W_N^{(i)}) \leq 2^{-N^\beta} \quad (17)$$

for $N$ sufficiently large. In particular, the bound (17) still holds if $\mathcal{A}_N$ is chosen as the polar set because by definition the polar set minimizes the sum in (17). Combining this fact with Proposition 1, it follows that

$$P_e(N,R,L) \leq 2^{-N^\beta}$$

for $N$ sufficiently large where $P_e(N,R,L)$ is defined as $P_e(N,L,\mathcal{A})$ for a polar set $\mathcal{A}$ with $|\mathcal{A}| = \lfloor NR \rfloor$. That is to say, using polar codes, with very high probability, the transmitted message will be on the list generated by the $L$-SCL decoder, provided that the rate is less than the symmetric capacity. Next, combining this fact with Proposition 2, it follows that

$$\bar{e}(N,R,L) \leq 2^{-N^\beta}$$

for $N$ sufficiently large. That is to say, also with very high probability, the transmitted message is the most likely message on the list. This completes the proof of Theorem 5.

*Remark*: One extreme case of Theorem 5 is when $L = 1$. In this case, it can be verified readily that the SCLG phase does all the jobs of an $L$-SCL decoder, the LMLS phase does nothing, and an $L$-SCL decoder degrades to an SC decoder. That is to say, SC decoders are special instances of $L$-SCL decoders.

*Remark*: Intuitively, an $L$-SCL decoder should behave better than an SC decoder. And the simulations in [2] have demonstrated this intuition. However, our proof cannot give out an exact order between $\bar{e}(N,R,L)$ and $\bar{e}(N,R,1)$ in general. This is because there is no exact inclusion relation between $\mathcal{Y}_L(m_1^N)$ and $\mathcal{Y}_1(m_1^N)$.

## III. SYMMETRIC CHANNELS

The main goal of this section is to prove the following result, which is a strengthened version of Theorem 5 for symmetric channels.

*Theorem 6*: For any symmetric B-DMC $\{W\}$, any fixed $R < I(W)$, $L \geq 1$, and $\beta < \frac{1}{2}$, every sequence of polar codes $(f_{G_N}, \varphi_{L-\text{SCL}})$ for $\{W\}$ with $N$ increasing to infinity, $|\mathcal{A}| = \lfloor NR \rfloor$, and $u_{\mathcal{A}^c}$ fixed arbitrarily satisfy

$$\bar{e}(N,L,\mathcal{A},u_{\mathcal{A}^c}) \leq 2^{-N^\beta}$$

for $N$ sufficiently large.

*Remark*: Recall that, by Theorem 5, there exists some choice of $u_{\mathcal{A}^c}$ such that the code $(f_{G_N}, \varphi_{L-\text{SCL}})$ can achieve the symmetric capacity $I(W)$. However, Theorem 6 tells us that, for symmetric channels, this is so for any choice of $u_{\mathcal{A}^c}$. Note also that, for symmetric channels, $I(W)$ equals the Shannon capacity.

*Proof of Theorem 6:* Consider an $N$-length $G_N$-coset code $(f_{G_N}, \varphi_{L-\text{SCL}})$ under $L$-SCL decoding for $\{W: \mathcal{X} \to \mathcal{Y}\}$ with respect to $\mathcal{A}$ and $u_{\mathcal{A}^c}$, only this time assuming that $\{W\}$ is a symmetric channel.

For any channel output $y_1^N \in \mathcal{Y}^N$, define a collection $\mathbb{L}_i(u_{\mathcal{A}_i^c}, y_1^N)$ of subsets of $\mathcal{X}^i$ and $\mathcal{C}_i(u_{\mathcal{A}_i^c}, y_1^N) \subseteq \mathcal{X}^i$ recursively in the order $i$ from 1 to $N$ where $\mathcal{A}_i^c \triangleq \mathcal{A}^c \cap \{1,\dots,i\}$ in the following way.

If $i \in \mathcal{A}^c$, then

$$\mathcal{C}_i(u_{\mathcal{A}_i^c}, y_1^N) \triangleq \{m_1^i: m_1^{i-1} \in \mathcal{C}_{i-1}(u_{\mathcal{A}_{i-1}^c}, y_1^N), m_i = u_i\}$$

and

$$\mathbb{L}_i(u_{\mathcal{A}_i^c}, y_1^N) \triangleq \{\{\hat{m}_1^i: \hat{m}_1^{i-1} \in \mathcal{L}_{i-1}, \hat{m}_i = u_i\}:$$

$$\mathcal{L}_{i-1} \in \mathbb{L}_{i-1}(u_{\mathcal{A}_{i-1}^c}, y_1^N)\}.$$

If $i \in \mathcal{A}$, let

$$\mathcal{B}_{\mathcal{C}_{i-1}} = \{m_1^i: m_1^{i-1} \in \mathcal{C}_{i-1}(u_{\mathcal{A}_{i-1}^c}, y_1^N), m_i \in \mathcal{X}\}.$$

For any $\mathcal{L}_{i-1} \in \mathbb{L}_{i-1}(u_{\mathcal{A}_{i-1}^c}, y_1^N)$, let

$$\mathcal{B}_{\mathcal{L}_{i-1}} = \{m_1^i: m_1^{i-1} \in \mathcal{L}_{i-1}, m_i \in \mathcal{X}\}$$

and

$$\mathcal{M}_{\mathcal{L}_{i-1}} = \{\hat{m}_1^i: (y_1^N, \hat{m}_1^i) \in g_{W_N^{(i)}}(\{y_1^N\} \times \mathcal{L}_{i-1}, \mathcal{X}, L)\}.$$

Let $\mathcal{T}_{\mathcal{L}_{i-1}} = \{t_1^i\}$ be such a set that either $t_1^i \in \mathcal{M}_{\mathcal{L}_{i-1}}$ and

$$W_N^{(i)}(y_1^N, t_1^{i-1}|t_i) = W_N^{(i)}(y_1^N, m_1^{i-1}|m_i)$$

for some $m_1^i \in \mathcal{B}_{\mathcal{L}_{i-1}} \setminus \mathcal{M}_{\mathcal{L}_{i-1}}$, or $t_1^i \in \mathcal{B}_{\mathcal{L}_{i-1}} \setminus \mathcal{M}_{\mathcal{L}_{i-1}}$ and

$$W_N^{(i)}(y_1^N, t_1^{i-1}|t_i) = W_N^{(i)}(y_1^N, m_1^{i-1}|m_i)$$

for some $m_1^i \in \mathcal{M}_{\mathcal{L}_{i-1}}$. Let

$$\mathcal{S}_{\mathcal{L}_{i-1}} = \mathcal{M}_{\mathcal{L}_{i-1}} \setminus \mathcal{T}_{\mathcal{L}_{i-1}}$$

and

$$\mathcal{S}_{\mathcal{L}_{i-1}}^c = \mathcal{B}_{\mathcal{L}_{i-1}} \setminus \mathcal{S}_{\mathcal{L}_{i-1}}.$$

Then

$$\mathbb{L}_i(u_{\mathcal{A}_i^c}, y_1^N) \triangleq \{\mathcal{L}_i: |\mathcal{L}_i| = |\mathcal{M}_{\mathcal{L}_{i-1}}|, \mathcal{L}_i \supseteq \mathcal{S}_{\mathcal{L}_{i-1}},$$

$$\mathcal{L}_i \setminus \mathcal{S}_{\mathcal{L}_{i-1}} \subseteq \mathcal{T}_{\mathcal{L}_{i-1}}, \mathcal{L}_{i-1} \in \mathbb{L}_{i-1}(u_{\mathcal{A}_{i-1}^c}, y_1^N)\}.$$

Define

$$\mathcal{E}_i(u_{\mathcal{A}_i^c}, y_1^N) \triangleq \mathcal{B}_{\mathcal{C}_{i-1}} \cap \left( \bigcup_{\mathcal{L}_{i-1} \in \mathbb{L}_{i-1}(u_{\mathcal{A}_{i-1}^c}, y_1^N)} \mathcal{S}_{\mathcal{L}_{i-1}}^c \right). \quad (18)$$

Then

$$\mathcal{C}_i(u_{\mathcal{A}_i^c}, y_1^N) \triangleq \mathcal{B}_{\mathcal{C}_{i-1}} \setminus \mathcal{E}_i(u_{\mathcal{A}_i^c}, y_1^N).$$

For every $i = 1, \dots, N$, from the definition of $\mathcal{C}_i(u_{\mathcal{A}_i^c}, y_1^N)$, it is obvious that $|\mathcal{C}_i(u_{\mathcal{A}_i^c}, y_1^N)| \leq L$. Hence, for each $i \in \mathcal{A}$, from (18), we know that $|\mathcal{E}_i(u_{\mathcal{A}_i^c}, y_1^N)| \leq 2L$ and for any $m_1^i \in \mathcal{E}_i(u_{\mathcal{A}_i^c}, y_1^N)$ there exists some $\mathcal{L}_{i-1} \in \mathbb{L}_{i-1}(u_{\mathcal{A}_{i-1}^c}, y_1^N)$ such that $m_1^i \in \mathcal{S}_{\mathcal{L}_{i-1}}^c$. Moreover, by the definition of $\mathcal{S}_{\mathcal{L}_{i-1}}^c$, some thought shows that there must exists some $\hat{m}_1^{i-1} \in \mathcal{L}_{i-1}$ such that

$$W_N^{(i)}(y_1^N, m_1^{i-1}|m_i)$$

$$\leq \sqrt{W_N^{(i)}(y_1^N, \hat{m}_1^{i-1}|0) W_N^{(i)}(y_1^N, \hat{m}_1^{i-1}|1)}$$

$$\leq \sum_{\substack{m_1^{i-1} \in \mathcal{X}^{i-1}: \\ m_{\mathcal{A}_{i-1}^c} = u_{\mathcal{A}_{i-1}^c}}} \sqrt{W_N^{(i)}(y_1^N, m_1^{i-1}|0) W_N^{(i)}(y_1^N, m_1^{i-1}|1)}.$$

Therefore, for any $i \in \mathcal{A}$, we have

$$\sum_{m_1^i \in \mathcal{E}_i(u_{\mathcal{A}_i^c}, y_1^N)} W_N^{(i)}(y_1^N, m_1^{i-1}|m_i)$$

$$\leq 2L \sum_{\substack{m_1^{i-1} \in \mathcal{X}^{i-1}: \\ m_{\mathcal{A}_{i-1}^c} = u_{\mathcal{A}_{i-1}^c}}} \sqrt{W_N^{(i)}(y_1^N, m_1^{i-1}|0) W_N^{(i)}(y_1^N, m_1^{i-1}|1)}. \quad (19)$$

Now, define

$$\mathcal{E} \triangleq \{(m_1^N, y_1^N) \in \mathcal{X}^N \times \mathcal{Y}^N : m_1^N \notin \widehat{\mathcal{M}}_1^N(m_{\mathcal{A}^c}, y_1^N)\} \quad (20)$$

and

$$\mathcal{E}_i \triangleq \{(m_1^N, y_1^N) \in \mathcal{X}^N \times \mathcal{Y}^N : m_1^i \in \mathcal{E}_i(m_{\mathcal{A}_i^c}, y_1^N)\}$$

for all $i \in \mathcal{A}$. Then, it is not difficult to show that

$$\mathcal{E} \subseteq \bigcup_{i \in \mathcal{A}} \mathcal{E}_i.$$

Moreover, from the definition of $\mathcal{E}_i$ and (2), it follows directly that

$$(m_1^N, y_1^N) \in \mathcal{E}_i \text{ implies } (a_1^N \oplus m_1^N, a_1^N G_N \cdot y_1^N) \in \mathcal{E}_i \quad (21)$$

for each $m_1^N, a_1^N \in \mathcal{X}^N, y_1^N \in \mathcal{Y}^N$. Thus, we have

$$\sum_{y_1^N : (m_1^N, y_1^N) \in \mathcal{E}_i} W_N(y_1^N | m_1^N)$$

$$= \frac{1}{2^N} \sum_{a_1^N \in \mathcal{X}^N} \sum_{y_1^N : (m_1^N, y_1^N) \in \mathcal{E}_i} W_N(a_1^N G_N \cdot y_1^N | a_1^N \oplus m_1^N)$$

$$= \sum_{(\widetilde{m}_1^N, y_1^N) \in \mathcal{E}_i} \frac{1}{2^N} W_N(y_1^N | \widetilde{m}_1^N)$$

$$= \sum_{y_1^N \in \mathcal{Y}^N} \sum_{\widetilde{m}_1^i \in \mathcal{X}^i : \widetilde{m}_1^i \in \mathcal{E}_i(\widetilde{m}_{\mathcal{A}_i^c}, y_1^N)} \frac{1}{2} W_N^{(i)}(y_1^N, \widetilde{m}_1^{i-1}|\widetilde{m}_i)$$

$$\leq L \sum_{y_1^N \in \mathcal{Y}^N, \widetilde{m}_1^{i-1} \in \mathcal{X}^{i-1}} \sqrt{W_N^{(i)}(y_1^N, \widetilde{m}_1^{i-1}|0) W_N^{(i)}(y_1^N, \widetilde{m}_1^{i-1}|1)}$$

$$= L Z(W_N^{(i)})$$

for all $m_1^N \in \mathcal{X}^N$ where the first equality is due to (1), the second equality is due to (21), and the inequality follows by (19).

Since $\mathcal{E} \subseteq \bigcup_{i \in \mathcal{A}} \mathcal{E}_i$, we obtain

$$\sum_{y_1^N : (m_1^N, y_1^N) \in \mathcal{E}} W_N(y_1^N | m_1^N) \leq L \sum_{i \in \mathcal{A}} Z(W_N^{(i)}) \quad (22)$$

for all $m_1^N \in \mathcal{X}^N$.

Next, as in the proof of Proposition 2, define

$$\mathcal{Y}_L(m_1^N) \triangleq \{y_1^N \in \mathcal{Y}^N : m_1^N \in \widehat{\mathcal{M}}_1^N(u_{\mathcal{A}^c}, y_1^N)\}$$

for all messages $m_1^N \in \mathcal{M}_{f_{G_N}}(u_{\mathcal{A}^c})$. Then, it holds by (20) and (22) that

$$W_N(\mathcal{Y}_L^c(m_1^N)|m_1^N) = \sum_{y_1^N : (m_1^N, y_1^N) \in \mathcal{E}} W_N(y_1^N|m_1^N)$$

$$\leq L \sum_{i \in \mathcal{A}} Z(W_N^{(i)}) \quad (23)$$

for all $m_1^N \in \mathcal{M}_{f_{G_N}}(u_{\mathcal{A}^c})$.

For any fixed $L \geq 1$, define

$$\mathcal{Y}(m_1^N) \triangleq \mathcal{Y}_1(m_1^N) \cap \mathcal{Y}_L(m_1^N)$$

for all messages $m_1^N \in \mathcal{M}_{f_{G_N}}(u_{\mathcal{A}^c})$. Then, by (23),

$$W_N(\mathcal{Y}^c(m_1^N)|m_1^N) = W_N(\mathcal{Y}_1^c(m_1^N) \cup \mathcal{Y}_L^c(m_1^N)|m_1^N)$$

$$\leq W_N(\mathcal{Y}_1^c(m_1^N)|m_1^N) + W_N(\mathcal{Y}_L^c(m_1^N)|m_1^N)$$

$$\leq 2L \sum_{i \in \mathcal{A}} Z(W_N^{(i)}) \quad (24)$$

for all $m_1^N \in \mathcal{M}_{f_{G_N}}(u_{\mathcal{A}^c})$.

In addition, as in the proof of Proposition 2, we know that the sets $\mathcal{Y}(m_1^N)$ are pairwise disjoint and the union of them is a subset of $\mathcal{Y}^N$ and

$$W_N(y_1^N|m_1^N) \leq W_N(y_1^N|\varphi_{L-\text{SCL}}(y_1^N)) \quad (25)$$

for all $y_1^N \in \mathcal{Y}(m_1^N)$.

Hence, it follows that

$$1 - \bar{e}(N, L, \mathcal{A}, u_{\mathcal{A}^c})$$

$$= \sum_{y_1^N \in \mathcal{Y}^N} \frac{1}{2^{|\mathcal{A}|}} W_N(y_1^N|\varphi_{L-\text{SCL}}(y_1^N))$$

$$\geq \sum_{m_1^N \in \mathcal{M}_{f_{G_N}}(u_{\mathcal{A}^c})} \sum_{y_1^N \in \mathcal{Y}(m_1^N)} \frac{1}{2^{|\mathcal{A}|}} W_N(y_1^N|\varphi_{L-\text{SCL}}(y_1^N))$$

$$\geq \sum_{m_1^N \in \mathcal{M}_{f_{G_N}}(u_{\mathcal{A}^c})} \sum_{y_1^N \in \mathcal{Y}(m_1^N)} \frac{1}{2^{|\mathcal{A}|}} W_N(y_1^N|m_1^N)$$

$$= \sum_{m_1^N \in \mathcal{M}_{f_{G_N}}(u_{\mathcal{A}^c})} \frac{1}{2^{|\mathcal{A}|}} W_N(\mathcal{Y}(m_1^N)|m_1^N)$$

$$\geq 1 - 2L \sum_{i \in \mathcal{A}} Z(W_N^{(i)})$$

where the first inequality is due to the fact that the sets $\mathcal{Y}(m_1^N)$ are pairwise disjoint and the union of them is a subset of $\mathcal{Y}^N$, the second inequality is due to (25), and the last inequality is due to (24).

This bound on $\bar{e}(N, L, \mathcal{A}, u_{\mathcal{A}^c})$ is independent of the frozen vector $u_{\mathcal{A}^c}$. Theorem 6 is now obtained by combining Theorem 2 with the above bound, as in the proof of Theorem 5. This completes the proof of Theorem 6.

## IV. PRECODED POLAR CODES

In this section, we formalize the codes in Improvement 2. In fact, we will formalize more general codes, i.e., precoded polar codes under SCL decoding, with our notation and prove that they achieve the symmetric capacity $I(W)$ of any given B-DMC $\{W\}$ asymptotically.

Given a B-DMC $\{W: \mathcal{X} \to \mathcal{Y}\}$, let $(f_{G_N}, \varphi_{L-\text{SCL}})$ be an $N$-length $G_N$-coset code under $L$-SCL decoding for $\{W\}$ with respect to $\mathcal{A}$ and $u_{\mathcal{A}^c}$ such that $|\mathcal{A}| = K$. We will use notation $\pi: \mathcal{X}^K \to \mathcal{X}^K$ to denote a permutation mapping on $\mathcal{X}^K$. For any such permutation $\pi$ and $k \leq K$, a $k$-to-$K$ precoder $\tilde{\pi}: \mathcal{X}^k \to \mathcal{X}^K$ induced by $\pi$ is defined as

$$\tilde{\pi}(m_1^k) \triangleq \pi(m_1^k, 0_{k+1}^K)$$

for all $m_1^k \in \mathcal{X}^k$.

*Definition 6*: An $N$-length precoded $G_N$-coset code $(f, \varphi)$ for $\{W\}$ with respect to the $G_N$-coset code $(f_{G_N}, \varphi_{L-\text{SCL}})$ and the precoder $\tilde{\pi}$ above is an $N$-length code for $\{W\}$ such that
i)     $\mathcal{M} = \mathcal{M}_f \triangleq \mathcal{X}^k$;
ii)    $\widehat{\mathcal{M}} \triangleq \mathcal{X}^k \cup \{*\}$;
iii)   $f(m_1^k) \triangleq f_{G_N}(\widetilde{m}_1^N) \in \mathcal{X}^N$ for all $m_1^k \in \mathcal{M}$ where $\widetilde{m}_{\mathcal{A}^c} = u_{\mathcal{A}^c}$ and $\widetilde{m}_\mathcal{A} = \tilde{\pi}(m_1^k)$;
iv)   The decoder $\varphi: \mathcal{Y}^N \to \widehat{\mathcal{M}}$ is defined as follows.
     For all $y_1^N \in \mathcal{Y}^N$, set

$$\mathcal{B}(y_1^N) \triangleq \widehat{\mathcal{M}}_1^N(u_{\mathcal{A}^c}, y_1^N)$$
$$\cap \{\widetilde{m}_1^N: \widetilde{m}_{\mathcal{A}^c} = u_{\mathcal{A}^c}, \widetilde{m}_\mathcal{A} = \tilde{\pi}(m_1^k), m_1^k \in \mathcal{M}\}.$$

If $\mathcal{B}(y_1^N) \neq \emptyset$, let

$$\{(y_1^N, \widehat{m}_1^N)\} = g_{W_N}(\{y_1^N\}, \mathcal{B}(y_1^N), 1),$$

then $\varphi(y_1^N) \triangleq \tilde{\pi}^{-1}(\widehat{m}_\mathcal{A})$.
If $\mathcal{B}(y_1^N) = \emptyset$, we can adopt at least three strategies.
Failure Strategy: The decoder fails with $\varphi(y_1^N) \triangleq *$. We will also use $\varphi_F$ to denote the decoder with this strategy.
Non-Failure Strategy: Let $\hat{v}_1^N = \varphi_{L-\text{SCL}}(y_1^N)$. Let $\widehat{m}_1^K = \pi^{-1}(\hat{v}_\mathcal{A})$. The decoder outputs $\varphi(y_1^N) \triangleq \widehat{m}_1^k$. We will also use $\varphi_{NF}$ to denote the decoder with this strategy.
Retransmission Strategy: The decoder requests the encoder to retransmit the message. We will also use $\varphi_R$ to denote the decoder with this strategy.
If $(f_{G_N}, \varphi_{L-\text{SCL}})$ is a polar code, we call $(f, \varphi)$ a precoded polar code.

With above formalization, we have the following result.

*Theorem 7*: The $N$-length precoded $G_N$-coset code $(f, \varphi)$ for $\{W\}$ defined above has the following properties.
The rate is $k/N$;
The error probabilities

$$e_{m_1^k}(W^N, f, \varphi) \leq e_{\widetilde{m}_1^N}(W^N, f_{G_N}, \varphi_{L-\text{SCL}}) \quad (26)$$

with $\widetilde{m}_{\mathcal{A}^c} = u_{\mathcal{A}^c}$ and $\widetilde{m}_\mathcal{A} = \tilde{\pi}(m_1^k)$,

$$e_{m_1^k}(W^N, f, \varphi_{NF}) \leq e_{m_1^k}(W^N, f, \varphi_F), \quad (27)$$

and

$$e_{m_1^k}(W^N, f, \varphi_R) \leq e_{m_1^k}(W^N, f, \varphi_F) \quad (28)$$

for all $m_1^k \in \mathcal{M}$.

Therefore, there exists a sequence of $N$-length precoded polar codes $(f, \varphi)$ that can achieve the symmetric capacity $I(W)$ asymptotically no matter which strategy has been adopted provided that

$$k/K \to 1 \text{ and } N^\beta - K + k \to \infty \quad (29)$$

as $N \to \infty$ for some $\beta < \frac{1}{2}$.

Moreover, if $\{W\}$ is a symmetric channel, any sequence of $N$-length precoded polar codes $(f, \varphi)$ can achieve the channel capacity $I(W)$ asymptotically no matter which strategy has been adopted provided that

$$k/K \to 1 \text{ and } N^\beta - K + k \to \infty$$

as $N \to \infty$ for some $\beta < \frac{1}{2}$.

*Proof.* The rate result is obvious.
For any $y_1^N \in \varphi_{L-\text{SCL}}^{-1}(\widetilde{m}_1^N)$ with $\widetilde{m}_{\mathcal{A}^c} = u_{\mathcal{A}^c}$ and $\widetilde{m}_\mathcal{A} = \tilde{\pi}(m_1^k)$, from the definition of $\varphi_{L-\text{SCL}}$, we have $\widetilde{m}_1^N \in \widehat{\mathcal{M}}_1^N(u_{\mathcal{A}^c}, y_1^N)$ and

$$\{(y_1^N, \widetilde{m}_1^N)\} = g_{W_N}(\{y_1^N\}, \widehat{\mathcal{M}}_1^N(u_{\mathcal{A}^c}, y_1^N), 1).$$

Hence, $\widetilde{m}_1^N \in \mathcal{B}(y_1^N)$ and $g_{W_N}(\{y_1^N\}, \mathcal{B}(y_1^N), 1) = \{(y_1^N, \widetilde{m}_1^N)\}$. Thus, $\varphi(y_1^N) = m_1^k$ and $y_1^N \in \varphi^{-1}(m_1^k)$. That is to say, $\varphi_{L-\text{SCL}}^{-1}(\widetilde{m}_1^N) \subseteq \varphi^{-1}(m_1^k)$ for all strategies, which proves (26).
Let

$$\widetilde{\mathcal{Y}} = \{y_1^N \in \mathcal{Y}^N: \mathcal{B}(y_1^N) \neq \emptyset\}.$$

In Failure Strategy,

$$\varphi_F^{-1}(*) = \widetilde{\mathcal{Y}}^c$$

and

$$e_{m_1^k}(W^N, f, \varphi_F) = 1 - W^N\left(\varphi_F^{-1}(m_1^k) | f(m_1^k)\right).$$

In Retransmission Strategy, as $\varphi_R(y_1^N) = \varphi_F(y_1^N)$ when $y_1^N \in \widetilde{\mathcal{Y}}$ and retransmission occurs when $y_1^N \in \widetilde{\mathcal{Y}}^c$, it follows that

$$e_{m_1^k}(W^N, f, \varphi_R) = 1 - \frac{W^N\left(\varphi_F^{-1}(m_1^k) | f(m_1^k)\right)}{W^N\left(\widetilde{\mathcal{Y}} | f(m_1^k)\right)},$$

which proves (28).
In Non-Failure Strategy, as $\varphi_{NF}(y_1^N) = \varphi_F(y_1^N)$ when $y_1^N \in \widetilde{\mathcal{Y}}$, it follows that

$$e_{m_1^k}(W^N, f, \varphi_{NF}) = 1 - W^N\left(\varphi_F^{-1}(m_1^k) | f(m_1^k)\right)$$
$$- \sum_{\substack{m_{k+1}^K \in \mathcal{X}^{K-k} \\ m_{k+1}^K \neq 0_{k+1}^K}} W^N\left(\varphi_{L-\text{SCL}}^{-1}\left(\pi(m_1^k, m_{k+1}^K)\right) \cap \widetilde{\mathcal{Y}}^c | f(m_1^k)\right),$$

which proves (27).

To prove the last claim, we note that, by Theorem 5, for the given $\beta$ and for any fixed $R < I(W)$, there exists a sequence of $N$-length polar codes $(f_{G_N}, \varphi_{L-\text{SCL}})$ such that

$$K = \lfloor NR \rfloor$$

and

$$\overline{e}(W^N, f_{G_N}, \varphi_{L-\text{SCL}}) \leq 2^{-N^\beta} \quad (30)$$

for $N$ sufficiently large. Now construct a sequence of $N$-length precoded polar codes $(f, \varphi)$ for $\{W\}$ with respect to these polar codes $(f_{G_N}, \varphi_{L-\text{SCL}})$ and some precoder $\tilde{\pi}$. Then, provided that $k/K \to 1$ and $N^\beta - K + k \to \infty$, we have the rate

$$\frac{k}{N} = \frac{k}{K}\frac{K}{N} \to R$$

and the average error probability

$$\begin{aligned}
\overline{e}(W^N, f, \varphi) &= \sum_{m_1^k \in \mathcal{M}_f} \frac{1}{2^k} e_{m_1^k}(W^N, f, \varphi) \\
&\leq \sum_{m_1^k \in \mathcal{M}_f} \frac{1}{2^k} e_{\tilde{m}_1^N}(W^N, f_{G_N}, \varphi_{L-\text{SCL}}) \\
&\leq 2^{K-k} \sum_{m_1^N \in \mathcal{M}_{f_{G_N}}} \frac{1}{2^K} e_{m_1^N}(W^N, f_{G_N}, \varphi_{L-\text{SCL}}) \\
&= 2^{K-k} \overline{e}(W^N, f_{G_N}, \varphi_{L-\text{SCL}}) \\
&\leq 2^{-(N^\beta - K + k)} \to 0
\end{aligned}$$

as $N \to \infty$ where the first inequality is due to (26) and the last inequality is due to (30).

The result for a symmetric channel follows similarly by using Theorem 6.

This completes the proof of Theorem 7.

*Remark*: One can verify that the CRC code used in [2] is one example of the precoder defined in this section. Moreover, as CRC codes have constant $K - k$, they satisfy the condition (29) in Theorem 7. Therefore, the precoded polar codes $(f, \varphi)$ for $\{W\}$ with respect to CRC precoders using Non-Failure Strategy, which indeed are the codes described in Section V of [2], can achieve the symmetric capacity $I(W)$. But the result of Theorem 7 is more general, which states that the precoded polar codes can achieve the symmetric capacity as long as the precoders satisfy the condition (29). Based on this, one may try to find out some other better precoders in future researches.

*Remark*: We know from (26) that the error probability of a message in $(f, \varphi)$ is no greater than that of the corresponding message in $(f_{G_N}, \varphi_{L-\text{SCL}})$, which has been confirmed by simulations with respect to CRC precoders in [2].

*Remark*: About the three strategies mentioned above, we know from (27) and (28) that both Non-Failure Strategy and Retransmission Strategy have better performance than Failure Strategy. But Retransmission Strategy can cause extra latency. In these senses, Non-Failure Strategy is the best.

## V. SOME EXTENSIONS AND CONCLUSION

First, the result of Theorem 5 is proven for binary DMC and $2 \times 2$ polarization matrix. In fact, polarization results have been proven for DMC over any finite field and $l \times l$ polarization matrix with any $l \geq 2$, see [3] for details. It is obvious that the definition of codes $(f_{G_N}, \varphi_{L-\text{SCL}})$ and the proof of Theorem 5 can be generalized to DMC over finite field and $l \times l$ polarization matrix straightforwardly. Thus, in fact, we have obtained the following more general result.

*Theorem 8*: For any DMC $\{W\}$ over any finite field and any $l \times l$ polarization matrix $G$ with any $l \geq 2$, let $E(G)$ be the exponent of $G$ and $N = l^n$ for $n \geq 0$. Then for any fixed $L \geq 1$, $R < I(W)$, and $\beta < E(G)$, there exists a sequence of $N$-length polar codes $(f_{G_N}, \varphi_{L-\text{SCL}})$ for $\{W\}$ with respect to $\mathcal{A}$ and $u_{\mathcal{A}^c}$ such that $|\mathcal{A}| = \lfloor NR \rfloor$ and $\overline{e}(W^N, f_{G_N}, \varphi_{L-\text{SCL}}) \leq 2^{-N^\beta}$ for $N$ sufficiently large.

Second, in Theorem 5, the list size $L$ is fixed. If we let $L_{N,i}$ be the list size at step $i$ of the SCL decoder of an $N$-length polar code where $i$ belongs to the polar set $\mathcal{A}$, i.e., we let the list size depend on the length $N$ and the step $i$, it is obvious that the proof of Theorem 5 is still valid provided that $L_{N,i}$ dose not decrease as $i$ increasing for each $N$.

*Remark*: One extreme case of this extension is when $L_{N,i} = 2^{|\mathcal{A}|}$. In this case, it can be verified readily that the SCLG phase does nothing, the LMLS phase does all the jobs of an SCL decoder, and an SCL decoder upgrades to an ML decoder. That is to say, ML decoders are special instances of SCL decoders. Hence, in fact, we have proved that polar codes under ML decoding can achieve the symmetric capacity, although it is trivial.

In conclusion, Arıkan proposed polar codes and proved that polar codes under SC decoding achieve the symmetric capacity $I(W)$ of any given B-DMC $\{W\}$ asymptotically in theory. But, in practice, simulation results show that there is a considerable performance gap between SC decoding and optimal ML decoding. That is to say, SC decoding is good in theory but not very good in practice. Then Tal and Vardy described SCL decoding and demonstrated in practice that SCL decoding can effectively bridge the proceeding performance gap between SC decoding and ML decoding. That is to say, SCL decoding is good in practice. But it was not known whether SCL decoding is good in theory or not, although SCL decoding includes SC decoding and ML decoding as special cases. In this paper, we have proven that polar codes under SCL decoding can achieve the symmetric capacity $I(W)$ of any given B-DMC $\{W\}$ asymptotically in theory. So, in fact, we have answered the proceeding question, i.e., SCL decoding is good in theory as well. All these together tell us that indeed SCL decoding can replace SC decoding as a common decoding method for polar codes both in practice and in theory.

**Zhuo Li** was born in Shannxi Province, China, in 1980. He received the B.S. degree (with highest honors) in mathematics, the M.S. degree in computer science, and the Ph.D. degree in information and communication engineering all from Xidian University, Xi'an, China, in 2002, 2005, and 2008, respectively.

During the years 2016 and 2017 he was a Visiting Scholar at the Texas A&M University, in College Station, TX. Since 2005 he has been with the school of telecommunication engineer of Xidian University, Xi'an, China, where he is presently a Professor. His research interests include information theory, coding theory, and quantum information and computing.

**Lijuan Xing** received the B.E. and Ph.D. degrees in information and communication engineering from Xidian University, Xi'an, China, in 2004 and 2008, respectively.

Since 2010 she has been with the school of telecommunication engineer of Xidian University, Xi'an, China, where she is presently an Associate Professor. Her research interests include communication systems, channel coding, and quantum information and computing.

**Ba-Zhong Shen** received his M.S. degree from Xidian University, China and his Ph.D. degree from Eindhoven University of Technology, the Netherlands.

From July 1993 to February 1996 he was with the Department of Electrical Engineering and Computer Science at Lehigh University, Bethlehem, PA, USA, becoming a Research Scientist in August 1993, and Adjunct Assistant Professor in January 1994. During 1996-1999, he was with Technology and Engineering of Quantum Corporation as principle design engineer. In 1999 he joined Broadcom Corporation, USA and later he became a technical director. Since May 2017 he has been with Xidian University, where he is a distinguished professor and dean of the school of telecommunication engineer.

He has more than 185 issued US patents. He contributes numerous industry standards including IEEE 802.3, IEEE 802.11, 3GPP LTE, ITU G.hn, MoCA, MPEG HEVC and DOCSIS. Many of his inventions have been adopted to the standards and built into the wire or wireless communication products. His interests include information theory, source and channel coding theory and technique, computing network and communication system.